\documentclass{elsart}
\usepackage{eurosym}
\usepackage{amssymb}
\usepackage{amsmath}
\usepackage{epsfig}
\usepackage{graphicx}

\journal{\textbf{ Lecture Notes in Physics, vol. 344, p.  85-98,
 Springer, Berlin (1989)}
\quad\quad\quad
\quad\quad\quad\quad\quad\quad\quad\quad\quad\quad\quad\quad\quad}
\input{tcilatex}
\begin{document}

\begin{frontmatter}
\title{Invariance Properties\\ of Inviscid Fluids of Grade n}

\author{Pierre Casal  \ \emph{\small and\ }}  \author[Francia] {Henri
Gouin} \ead{(Corresponding author)  henri.gouin@univ-cezanne.fr }
\address[Francia]{University of Aix-Marseille \& C.N.R.S. \ U.M.R.
6181\\
 Av. Escadrille Normandie-Niemen, Box 322, 13397 Marseille Cedex 20, France.  }

{\small \texttt{\emph{Revisited version in memory of Professor Pierre Casal}}}
\begin{abstract}
    Fluids of grade n are continuous media  in dynamic changes
    of phases  avoiding the surfaces of discontinuity and
    representing the capillary layers in liquid-vapour interfaces.
    We recall the thermodynamic form of the equation of motion for
    inviscid fluids of grade n [1]. \emph{First integrals and theorems
    of circulation are deduced. A general classification of flows is
    proposed.}
\end{abstract}

\end{frontmatter}

\section{Introduction}

In continuum mechanics, media of first gradient order cannot represent
fluids with strong density variations. Material surfaces need their own
characteristic behaviour and energy properties [2]. \newline
Fluids of grade $n$ ($n$ is any integer greater than \emph{1}) are
continuous media with an internal energy per unit mass $\varepsilon$ which
is a function of the entropy $s$, the density $\rho $ and spatial gradients
up to the $n-\emph{1}$ order:
\begin{equation*}
\varepsilon =\varepsilon \left(s,\mathrm{\ \func{grad}\, }s,...,({\func{grad}}%
)^{(n-1)}s,\rho ,\mathrm{\func{grad}}\,\rho ,...,(\mathrm{\func{grad}}%
)^{(n-1)}\rho\right)
\end{equation*}
D.J. Korteweg pointed out the advantage of fluids of grade n with respect to
the density and removed the discontinuity surfaces representing the
capillary layers in liquid-vapour interfaces [3]. \newline
We improve the model accuracy by taking into account the successive
gradients of density. Recently, this approach was used in the case of
dynamic changes of phase [4]. Till now, the thermodynamic part of the fluid
has been neglected. Due to the fact it is not possible to consider a virtual
displacement of temperature, it is necessary to use the entropy by the way
of the internal energy. \newline
The equation of motion with both strong gradients of density and entropy is
obtained. One can accord importance to the fact that the thermodynamic form
of the equation of motion is independent of the order $n$. Flows with strong
variations of temperature such as those associated with combustion phenomena
or non-isothermal interfaces may be considered.

The conservative flows of classical perfect fluids are only a mathematical
model. It is also the case for fluids of grade $n$. The study of their
structure is necessary. The conclusions for fluids of grade $n$ are of the
same kind as for compressible fluid [5]: it is possible to obtain a Clebsh
transformation of the motion equations (\emph{potential equations}) and many
invariance properties and first integrals. The potential equations may be
used to classify motions in the same manner as in the case of perfect fluids
[6,7]. \newline
The simplest case is for fluids of grad 2 (or thermocapillary fluids): the
internal energy is a function only of the density, the entropy and their
first spatial derivatives
\begin{equation*}
\varepsilon =\varepsilon (s,\mathrm{\func{grad}}\,s,\rho ,\mathrm{\func{grad}%
}\,\rho)
\end{equation*}
and $n>2 $ adds little except for complications in the momentum balance
equation. If we consider flows through liquid-vapour interfaces and use
convenient physical units, the equations of thermocapillary fluids are able
to study non-isothermal motions and yield a possible interpretation of film
boiling phenomena [8,9].

\section{Thermodynamic form of the equation of motion for inviscid fluids of
grade n}

The principle of virtual works allows to obtain the equation of motion. In
the case of a perfect fluid motion, it is written by means of the Hamilton
principle [5,6,10]. \newline
The variations of particle motions are deduced from families of virtual
motions of the fluid in the form [5]
\begin{equation*}
\mathbf{X}=\Psi (\mathbf{x},t,\alpha )
\end{equation*}%
where $\alpha $ is a parameter defined in a neighbourhood of zero. The real
motion corresponds to $\alpha =0$. \newline
A particle is represented in Lagrange coordinates by $\mathbf{X}\
(X^{1},X^{2},X^{3})$ as a position in a reference space\ $\mathcal{D}_{o}$%
\thinspace . At time $t$ its position is given in $\mathcal{D}_{t}$ by the
Eulerian representation $\mathbf{x}\ (x^{1},x^{2},x^{3})$ (see Appendix).
The virtual displacements associated with a variation of the real motion can
be written
\begin{equation*}
\delta \mathbf{X}=\left. \frac{{\partial \Psi }}{{\partial \alpha }}(\mathbf{%
x},t,\alpha )\right\vert _{\alpha =0}
\end{equation*}%
The variation is dual with Serrin's [5]: the two variations are
mathematically equivalent [11]. The Lagrangian of a fluid of $\,\mathrm{%
\func{grad}}\,n$ is defined as
\begin{equation*}
L=\rho \,\left( \frac{1}{2}\,\mathbf{V}^{\ast }\mathbf{V}-\varepsilon
-\Omega \right) ,
\end{equation*}%
where $\mathbf{V}$ denotes the velocity, $\Omega $ denotes the extraneous
force potential defined in $\mathcal{D}_{t}$ and $^{\ast }$ is the
transposition in $\mathcal{D}_{t}$ such that $\mathbf{V}^{\ast }\mathbf{V}$
is the scalar product of $\mathbf{V}$ by $\mathbf{V}$. Between times $t_{1}$
and time $t_{2}$, the Hamiltonian action is
\begin{equation*}
a=\int\nolimits_{t_{1}}^{t_{2}}\int\nolimits_{D_{t}}L\ dvdt
\end{equation*}%
The density verifies the balance equation
\begin{equation*}
\rho \,\det F=\rho _{o}(\mathbf{X}),
\end{equation*}%
where $\rho _{o}$ is defined in $\mathcal{D}_{o}$ and $F$ is the deformation
gradient. We deduce
\begin{equation}
\frac{d\rho }{dt}+\rho \ \mathrm{{{\func{div}\ }}}\mathbf{V}=0
\label{density}
\end{equation}%
The entropy variation is the sum of a variation associated with the virtual
motion and another one, $\delta _{1}s$, related to the particle [9]
\begin{equation*}
\delta s=\left( \frac{\partial s}{\partial \mathbf{X}}\right) \,\delta
\mathbf{X}+\delta _{1}s,
\end{equation*}%
where $\delta _{1}s$ and $\delta \mathbf{X}$ are independent.To obtain the
equation of motion, we consider the conservative case corresponding to $%
\delta _{1}s$ $=0$. \newline
Classical methods of variational calculus provide the Hamiltonian action
variation
\begin{equation*}
\delta a=0
\end{equation*}%
A calculus as in [1] yields
\begin{equation*}
\delta a=\int_{t_{1}}^{t_{2}}\int_{\mathcal{D}_{t}}\rho _{o}\left[ \frac{d}{%
dt}(\mathbf{V}^{\ast }F)-\theta \,\mathrm{{{grad}_{o}^{\ast }}}\,s-\mathrm{{{%
grad}_{o}^{\ast }}}\,m\right] \delta \mathbf{X}\,dv_{o}dt,
\end{equation*}%
where\newline
\begin{equation}
\left\{
\begin{array}{l}
\displaystyle m=\frac{1}{2}\,\mathbf{V}^{\ast }\mathbf{V}-h-\Omega \\
\displaystyle h=\varepsilon +\frac{p}{\rho } \\
\displaystyle p=\rho ^{2}\varepsilon _{\rho }^{\prime }+\rho
\sum_{k=1}^{n-1}(-1)^{k}\mathrm{div}^{(k)}\left( \rho \frac{\partial
\varepsilon }{\partial \left( {\func{grad}}^{(k)}\rho \right) }\right) \\
\displaystyle\theta =\ \varepsilon _{s}^{\prime }+\frac{1}{\rho }%
\sum_{k=1}^{n-1}(-1)^{k}\mathrm{div}^{(k)}\left( \rho \frac{\partial
\varepsilon }{\partial \left( {\func{grad}}^{(k)}s\right) }\right)
\end{array}%
\right.  \label{quantities}
\end{equation}%
We have taken into account the relation $\delta \rho =\rho \,\mathrm{{{\func{%
div_o}}}}(\delta \mathbf{X})+\dfrac{1}{\det F}\dfrac{\partial \rho _{o}}{%
\partial \mathbf{X}}\,\delta \mathbf{X}$ and $\delta \mathbf{V}=-F\,\dfrac{%
d(\delta \mathbf{X})}{dt}$; $\func{div}^{(k)}$ and $\mathrm{{{\func{grad}}}}%
^{(k)}$ are the divergence and the gradient operators in $\mathcal{D}_{t}$
reiterated $k$-times; $\mathrm{{{div}_{o}}}$ and $\mathrm{{{grad}_{o}}}$ are
the divergence and the gradient operators in $\mathcal{D}_{o}$. \newline
Obviously, $p,\theta $ and $h$ have respectively the dimension of a
pressure, a temperature and a specific enthalpy. We call these quantities
\emph{the pressure, temperature and enthalpy of the fluid of grade n}%
\thinspace .

The Hamilton principle yields:

\emph{%
\centerline{{ For any displacement} $\
\delta \mathbf{X}\, $ null on the edge of\ \ $\mathcal{D}_{o}$,\ $\delta a=0$ . }%
}

We get the equation of motion
\begin{equation*}
\frac{{d(\mathbf{V}^{\ast }F)}}{{dt}}-\theta \,{\mathrm{{{grad}_{o}^{\ast }}%
\,s-{grad}_{o}^{\ast }\,m=0}}
\end{equation*}%
Let us note that $({\mathbf{\Gamma }}^{\ast }+\mathbf{V}^{\ast }\displaystyle%
\frac{{\partial \mathbf{V}}}{{\partial \mathbf{x}}})F=\frac{{d(\mathbf{V}%
^{\ast }F)}}{dt}$, we obtain
\begin{equation}
{\mathbf{\Gamma }}=\theta \ {\mathrm{grad}}\,s-{\mathrm{grad}}(h+\Omega )
\label{motion1}
\end{equation}%
In case of inviscid thermocapillary fluids the dissipative function is null
and the equations of motion and energy imply [12]:
\begin{equation}
\rho \,\theta \,\frac{ds}{dt}+\mathrm{div}\,\mathbf{q}-r=0  \label{Planck}
\end{equation}%
where $\mathbf{q}$ is the heat flux vector and $r$ the heat supply. This
result is extended to the fluids of grade n [13]. \newline
A constitutive equation must be added to these equations, which yields the
behaviour of the heat flux vector $\mathbf{q}$. For example, the Fourier law
is
\begin{equation}
\begin{array}{lllll}
\mathbf{q}=-k\,\mathrm{{{grad}\,\theta }} &  & \mathrm{and} &  & k\geq 0%
\end{array}
\label{Fourier}
\end{equation}%
The heat supply is assumed to be given in similar way than the extraneous
force potential. From the heat conduction inequality, we obtain the
Clausius-Duhem inequality and the fluids of grade $n$ are compatible with
the second law of thermodynamics [14,15]
\begin{equation*}
\rho \,\frac{ds}{dt}+\mathrm{{{div}\left( \frac{\mathbf{q}}{\theta }\right) -%
\frac{r}{\theta }\geq 0}}
\end{equation*}%
Consequently, the equations of motions of inviscid fluids of grade n verify
\begin{equation}
\left\{
\begin{array}{l}
{\mathbf{\Gamma }}=\theta \,{\mathrm{grad}}\,s-{\mathrm{grad}}(h+\Omega ) \\
\dfrac{d\rho }{dt}+\rho \,\mathrm{{{div} \mathbf{V} =0}} \\
\rho \,\theta
\,\displaystyle\frac{ds}{dt}+\mathrm{{{div}\,\mathbf{q}-r=0}}
\\
\begin{array}{lllll}
\mathbf{q}=-k\,\mathrm{{{grad}\,\theta }} &  & \mathrm{and} &  & k\geq 0 %
\end{array}%
\end{array}%
\right.  \label{process}
\end{equation}%
The thermodynamic form of the motion equation of inviscid fluids is
equivalent to the classic balance equation. The thermodynamic form is the
same for all fluids (independently of the order $n$). Nevertheless, the
complexity of balance equations increases with $n$. In fact, only the
expressions for the temperature and the enthalpy become more complex and we
have to modify the definitions of these quantities\footnote{%
The definitions are given by system (\ref{quantities}).}.

\section{Thermodynamic hypothesis on the motion}

If we study the invariance properties of the motions, we must assume that
the heat supply allows a behaviour expressing that a thermodynamic quantity $%
T$ has a zero material derivative or - with a stronger assumption - is
constant everywhere in the flow. Consequently equations (\ref{Planck}) and (%
\ref{Fourier}) are replaced by one of the two conditions
\begin{equation}
\displaystyle\frac{{dT}}{dt}=0 \qquad\qquad \mathrm{or}\qquad\qquad T=C^{ste}
\label{temperature}
\end{equation}%
For example, we will consider the case of an isentropic motion ($T=s$).
Then,
\begin{equation*}
\displaystyle\frac{{ds}}{dt}=0
\end{equation*}
For the case of an isothermal motion ($T=\theta $). Then,
\begin{equation*}
\theta =C^{ste}
\end{equation*}
They are two limit cases: one is associated with fast motions, the other
with slow motions\footnote{%
The Euler equation represents the equation of motion for classical inviscid
fluids. Such motions can be obtained with non-negligible heat flux and heat
supply. They correspond to physical situations. As a matter of fact, in the
non-dimensional equation of motion for viscous fluids [16,17], it is often
justified to neglect the viscous terms when, with respect to other physical
quantities, the Reynolds number is large. Nevertheless, for large
temperature gradients and with a Prandlt number of order one, it is not
possible to remove the terms of thermal conductivity. Such a case arises
when we study the "thermal" boundary layer [18,19] and slow motions of
natural convection in fluid with gravity and heat sources associated with
large differences of temperature [20].}. Then, the fluid motions satisfy the
system consisting of equations (\ref{density}), (\ref{motion1}) and (\ref%
{temperature}).

We consider the more general case when $T$ is a differentiable function of $%
\theta$ and $s$ such that $\displaystyle\frac{{\partial T}}{{\partial \theta
}}\neq 0$. Consequently, we denote by $H(T,s)$ a differentiable function of
the variables $T$ and $s$ such that \ $\displaystyle\frac{{\partial H}}{{%
\partial s}}=\theta (T,s)$.\newline
The function $H$ is defined, to an unknown function of $T$, by the
differential form
\begin{equation*}
dH=\theta\, ds-u\,dT
\end{equation*}%
The two variables $u$ and $T$ can be used instead of $\theta $ and
$s$ to study the general case by transforming Eq. (\ref{motion1})
following the method of Lemma 1. Let us denote $g=h-H$, Eq.
(\ref{motion1}) can be written:
\begin{equation}
\mathbf{\Gamma }=u\,\mathrm{{{grad}\,T-{grad}\,(g+\Omega )}}  \label{motion2}
\end{equation}%
Let us note that if $T=\theta $, $g$ is the free enthalpy
($g=h-\theta s$) and $u=-s$, and the motions satisfy System
 ((\ref{density}),(\ref{temperature}),(\ref{motion2})).\newline First integrals are obtained by the
research of invariant scalars as the state variables of the fluid
(the density, the entropy, the velocity, with a zero material
derivative) and expression (\ref{motion2}) allows us to make a
single classification for the fluids.

The momentum equation expresses the balance of forces (inertial forces, body
forces and stresses).

Let us note that System
((\ref{density}),(\ref{temperature}),(\ref{motion2})) is
not in balance form, but expresses the fluid evolution: the relation $%
\displaystyle\frac{dT}{dt}=0$ implies  $T$ is constant along trajectory and
the equation of continuity $\displaystyle\frac{d\rho }{dt}=-\rho\, \mathrm{{{%
div}\mathbf{V}}} $ yields the density variation. It will be shown that
equations (\ref{motion1}) or (\ref{motion2}) describe the evolution of the
velocity field.

\section{Material and convective derivatives}

The material derivative of any tensorial quantity expressed in the physical
space is the derivative with respect to time when one follows the particle
along the motion.\newline
The tensorial quantity has an image in the reference space. The image in the
physical space of the derivative with respect to time of the tensorial
quantity in the reference space is the value of the convective derivative.
The convective derivative is denoted by $d_{c}$\, (see Appendix).

The material and the convective derivatives of a scalar are equal. A tensor
is \emph{convected by the flow} when its convective derivative is null.

If we transpose Eq. (\ref{motion2}), we get the convective derivative of the
velocity covector $\mathbf{V}^{\ast}$ in the form
\begin{equation}
d_{c}\mathbf{V}^{\ast }=\mathrm{{{grad}^{\ast }\left(\frac{1}{2}\, \mathbf{V}%
^{2}-g-\Omega\right)+u \ {grad}^{\ast }\,T}} .  \label{motion3}
\end{equation}%
Eq. (\ref{motion3}) is equivalent to Eq. (\ref{motion2}) and yields \emph{%
the Helmholtz-Kelvin theorem in its most general form}:\newline
Let us consider an arbitrary fluid curve $\mathcal{C}$ such that $I(t)=%
\displaystyle \int_{\mathcal{C}}\mathbf{V}^{\ast }d\mathbf{x}$ is the
circulation of the velocity field. The circulation of the convective
derivative of the velocity field is
\begin{equation*}
\frac{{dI}}{{dt}}=\displaystyle \int_{\mathcal{C}}d_{c}\mathbf{V}^{\ast }d%
\mathbf{x}=\int_{\mathcal{C}}d \left(\frac{1}{2}\,\mathbf{V}^{2}-g-\Omega
\right)+\int_{\mathcal{C}}u\,dT
\end{equation*}%
If $\mathcal{C}$ is a closed curve, the integral $\displaystyle \int_{%
\mathcal{C}}d \left(\frac{1}{2}\,\mathbf{V}^{2}-g-\Omega \right)$ is null.
Moreover, if $\mathcal{C}$ is located on a $T$-surface (surfaces $T=C^{ste}$
are fluid surfaces and this assumption is compatible with the fact that $%
\mathcal{C}$ is a fluid curve), the second integral is null and \emph{$I$ is
constant along the motion.}

When the entropy is constant all over the fluid $(T=s=C^{ste})$, the
so-called Kelvin theorem corresponds to the special case of \textit{%
homentropic} motions: \emph{the circulation of the velocity field on any
closed curve is constant}.

Due to Noether's theorem, it is known that any conservation law can be
represented by a group of invariance. It has been shown the conservation law
expressed by Helmholtz-Kelvin theorems associated with isentropic fluid
curves corresponds to the group of the permutations of the particles of
equal entropy [21, 22]. This group keeps the equations of motion invariant
for both a classical perfect fluid and a perfect fluid of grade $n$. We
propose to define a \emph{general perfect fluid} as a fluid invariant by
this group or as a continuous medium whose motions are submitted to
Helmholtz-Kelvin theorems.

\section{Potential equations}

Let $\varphi $ and $\psi $ be two scalar fields such that the material
derivatives verify
\begin{equation*}
\begin{array}{lllll}
\displaystyle\frac{{d\varphi }}{{dt}}=\frac{1}{2}\,\mathbf{V}^{2}-g-\Omega &
& \mathrm{and} &  & \displaystyle\frac{{d\psi }}{dt}=u .%
\end{array}%
\end{equation*}%
Equation (\ref{motion3}) is equivalent to
\begin{equation*}
d_{c}\mathbf{V}^{\ast }=d_{c}(\mathrm{{{grad}^{\ast }\,\varphi +\psi \,{grad}%
\,^{\ast }T)}}.
\end{equation*}%
The covector $\mathbf{V}^{\ast }$ and the covector $\mathbf{W}^{\ast }=%
\mathrm{{{grad}^{\ast }\,\varphi +\psi \,{grad}^{\ast }\,T}}$ have the same
convective derivative. The difference $\mathbf{(V-W)}^{\ast }$ is a covector
convected by the flow, which leads to the system
\begin{equation}
\left\{
\begin{array}{c}
\mathbf{V}=\mathrm{{{grad}\,\varphi +\psi \,{grad}\,T+\mathbf{C}}}, \\
\\
\displaystyle\quad\frac{{dT}}{{dt}}=0,\quad \displaystyle\frac{{d\varphi }}{%
dt}=\frac{1}{2}\,\mathbf{V}^{2}-g-\Omega ,\quad \displaystyle\frac{{d\psi }}{%
dt}=u,\quad d_{c}\mathbf{C}^{\ast }=0.%
\end{array}%
\right.  \label{equapot}
\end{equation}%
System ((\ref{density}),(\ref{temperature}),(\ref{motion2})) is
equivalent to system ((\ref{density}),(\ref{equapot})).\newline
In this case, the generalized Helmholtz-Kelvin theorem is reduced to \emph{%
the circulation of $\ \mathbf{C}$ along any fluid curve is constant.}

Due to the Stokes formulae, the vector $\mathbf{U}=\displaystyle\frac{1}{%
\rho }\,\mathrm{rot}\,\mathbf{C}$ is convected by the flow ($d_{c}\mathbf{U}%
=0$) i.e. :
\begin{equation}
\displaystyle\frac{d\mathbf{U}}{dt}-\frac{\partial \mathbf{V}}{\partial x}\
\mathbf{U}=0  \label{motion4}
\end{equation}%
Due to the fact that $\mathbf{C}=\gamma \,{grad}\,\nu $, where $\gamma $ and
$\nu $ are two scalars \emph{convected by} the flow and $\mathrm{{{\ grad}%
\,\varphi ,}}$ $\mathrm{grad}\,\mathrm{\psi ,}$ $\mathrm{grad}\,\mathrm{\
\nu }$ are three independent vectors (see Lemma 2 and Theorem 3), the system
(\ref{equapot}) leads to the system \emph{of potential equations }
\begin{equation}
\left\{
\begin{array}{c}
\mathbf{V}=\mathrm{{{grad}\,\varphi +\psi \,{grad}\,T+\gamma \,{grad}\,\nu }}%
, \\
\\
\displaystyle\quad\frac{{dT}}{{dt}}=0,\quad \displaystyle\frac{{d\varphi }}{%
dt}=\frac{1}{2}\,\mathbf{V}^{2}-g-\Omega ,\quad \displaystyle\frac{{d\psi }}{%
dt}=u,\quad \displaystyle\frac{{d\gamma }}{dt}=0,\quad \displaystyle\frac{{%
d\nu }}{dt}=0.%
\end{array}%
\right.  \label{equapot1}
\end{equation}%
In the general case, the velocity field is associated with five potentials; $%
T$ is one of them\footnote{%
The velocity field depends also on the density which is not considered as a
potential.}.

\section{ Classification of motions}

As for $T$, the \emph{potentials} $\gamma $ and $\nu $ have a zero material
derivative. They are first integrals of the motion [23]. \newline
These potentials have constant values on each trajectory; when they are
constant in a fluid domain, the values remain constant in time-space.
\newline
In the domain convected by the flow, the fluid velocity is expressed by
fewer potentials characterizing the kinematic properties of the motion. In
an adjoining fluid domain, the motion may be different. Then, the flow is
separated into different domains that move, remain adjacent and keep their
own characteristics.

Due to the absence of viscosity, in hydrodynamics of incompressible perfect
fluids the flow is divided into vortical and irrotational domains that do
not mix. The following study shows that, for all inviscid fluids of any
grade (compressible or not), the vortical domain may be divided into several
parts of different kinds\footnote{%
System (\ref{equapot1}) assumes the flow to be without discontinuity. A
shock wave transforms the nature of flow [11, 24].} \newline
First integrals are tensorial quantities of different nature. For a
non-scalar tensor, the first integral is a tensor convected by the flow (or
with a zero convective derivative) i.e. has a constant tensorial image in a
reference space.

\subsection{Oligotropic motions}

When the covector $\mathbf{C}^{\ast }$ is zero (i.e. $\gamma =0$ or $\nu
=C^{ste}$. ), the velocity field has a particular kinematic type. We have
the following properties:

\quad (a) The velocity field can be expressed by the use of three potentials
and the equations of motion are
\begin{equation}
\left\{
\begin{array}{c}
\mathbf{V}=\mathrm{grad}\,\varphi +\psi \,\mathrm{grad}\,T , \\
\\
\begin{array}{ccccc}
\displaystyle\quad\frac{dT}{dt}=0, &  & \displaystyle\frac{{d\varphi }}{dt}=%
\frac{1}{2}\,\mathbf{V}^{2}-g-\Omega , &  & \displaystyle\frac{{d\psi }}{dt}%
=u.%
\end{array}%
\end{array}%
\right.  \label{oligotropic}
\end{equation}

\quad (b) From System (\ref{oligotropic}) we get
\begin{equation*}
\mathrm{rot}\,\mathbf{V}=\mathrm{grad}\,\psi \wedge \mathrm{grad}\,T,
\end{equation*}%
then
\begin{equation}
\mathrm{{{grad}^{\ast }}}\,T\ \mathrm{rot}\,\mathbf{V}=0
\label{oligotropic2}
\end{equation}%
Consequently, \emph{the vortex lines are located on the surfaces} $T=C^{ste}$%
.

Property (\ref{oligotropic2}) is equivalent to System (\ref{oligotropic}).
The motions, we call \emph{oligotropic} motions, are intermediate between
irrotational and general motions. In System (\ref{oligotropic}), the
expression of the velocity is independent of time and the motion remains
oligotropic at any moment.

\subsection{Homentropic motions}

The thermodynamic condition is $T=C^{ste}$ (which is stronger than $%
\displaystyle\frac{dT}{dt}=0$) and the velocity field is of another
particular kind. When $T=s$ , the motion is not only isentropic but also
homentropic. We have the following properties:

\quad (a) The velocity field can be expressed with the use of three
potentials and the equations of motion are
\begin{equation*}
\left\{
\begin{array}{c}
\mathbf{V}=\func{grad}\varphi +\psi \func{grad}\nu, \\
\\
\begin{array}{lllllll}
\displaystyle\frac{dT}{dt}=0, &  & \displaystyle\frac{{d\varphi }}{dt}=\frac{%
1}{2}\,\mathbf{V}^{2}-g-\Omega , &  & \displaystyle\frac{{d\gamma }}{dt}=0,
&  & \displaystyle\frac{{d\nu }}{dt}=0 .%
\end{array}%
\end{array}%
\right.
\end{equation*}%
In fact the Clebsch representation [16] is practically worthless because the
curves $\gamma =C^{ste}$ and $\nu =C^{ste}$ are the vortical lines whose
complexity is well known [25]. It is easier to write the motion equations in
an equivalent form:
\begin{equation*}
\left\{
\begin{array}{c}
\mathbf{V}=\func{grad}\varphi +\mathbf{C}, \\
\\
\begin{array}{ccccc}
\displaystyle\frac{dT}{dt}=0, &  & \displaystyle\frac{{d\varphi }}{dt}=\frac{%
1}{2}\,\mathbf{V}^{2}-g-\Omega , &  & d_{c}\mathbf{C}^{\ast }=0 .%
\end{array}%
\end{array}%
\right.
\end{equation*}

\quad (b) Vector $\displaystyle \frac{1}{\rho} \ \mathrm{rot}\,\mathbf{V}$
is a convected vector
\begin{equation}
d_{c}\left( \frac{1}{\rho }\,\mathrm{rot}\,\mathbf{V}\right)=0\qquad
\label{homent}
\end{equation}%
When a thermodynamic potential $T$ is constant in the flow, the Cauchy
theorem for homentropic motions is generalized and Eq. (\ref{homent}) is
characteristic of these motions.

\subsection{Irrotational motion}

If an oligotropic motion is a motion where $T$ is constant, the velocity
field derives from only one potential
\begin{equation*}
\left\{
\begin{array}{l}
\quad\mathbf{V}=\mathrm{grad}\,\varphi , \\
\\
\displaystyle\quad\frac{{d\varphi }}{dt}=\,\frac{1}{2}\,\mathbf{V}%
^{2}-g-\Omega%
\end{array}%
\right.
\end{equation*}%
At every time, $\mathrm{rot}\,\mathbf{V}=\,0$ \footnote{%
When at one moment $\mathrm{rot}\,\mathbf{V}=0$, this property is
not necessarily verified at every time as in hydrodynamics of
compressible fluids.}. This is the case of irrotational motions: the
velocity whose the evolution corresponds to the equation of
Bernoulli is the gradient of a single potential.

\begin{figure}[tbp]
\begin{center}
\epsfig{file={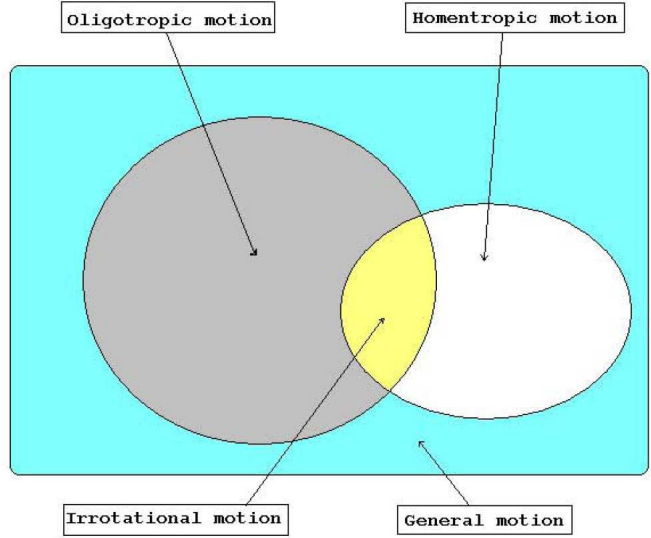}, width = 0.65\linewidth}
\end{center}
\caption{For perfect fluid motions, the fluid domains are
distributed in four parts. Convected by the flow, the domains become
deformed but do not mix. } \label{fig1}
\end{figure}

\subsection{Permanent motions}

For a permanent motion, Equation (\ref{motion2}) may be written in a
generalized Crocco-Vaszonyi form
\begin{equation}
\mathrm{rot}\mathbf{V}\wedge \mathbf{V}=u\,\mathrm{grad}\,T-\mathrm{grad}\,G
\label{crocco}
\end{equation}%
where $G=\displaystyle\frac{1}{2}\,\mathbf{V}^{2}+g+\Omega $.\newline
If $T=s$, $g$ is the specific enthalpy and $G$ the Lamb function [16]
(surfaces where $G=C^{ste}$ are Lamb surfaces). Consequently,
\begin{equation*}
\displaystyle\frac{dT}{dt}=0\quad \Leftrightarrow \quad \displaystyle\frac{dG%
}{dt}=0
\end{equation*}%
$G$ is a scalar first integral of the motion and we use $G$ for function $%
\nu $. Potential equations for general permanent flows are
\begin{equation*}
\left\{
\begin{array}{c}
\mathbf{V}=\func{grad}\varphi _{1}+\psi _{1}\func{grad}T+\gamma _{1}\func{%
grad}G, \\
\\
\begin{array}{lllllllll}
\displaystyle\quad\frac{{d\varphi }_{1}}{dt}=\frac{1}{2}\,\mathbf{V}%
^{2}-g-\Omega , &  & \displaystyle\frac{{d\psi }}{dt}=u, &  & \displaystyle%
\frac{dT}{dt}=0, &  & \displaystyle\frac{d\gamma _{1}}{dt}=0, &  & %
\displaystyle\frac{dG}{dt}=0.%
\end{array}%
\end{array}%
\right.
\end{equation*}%
Let us note: $\displaystyle\frac{dT}{dt}=0$ and $\displaystyle\frac{dG}{dt}=0
$ $\ \Rightarrow \ \displaystyle\frac{\partial \varphi _{1}}{\partial t}=-G$%
; if $\varphi =\varphi _{1}+tG$, $\varphi $ is independent of time. Let us
denote $\gamma _{1}=\gamma +t$ and $\psi _{1}=\psi $, the equations of
motion can be written with potentials independent of time
\begin{equation}
\left\{
\begin{array}{c}
\mathbf{V}=\func{grad}\varphi +\psi \func{grad}T+\gamma \func{grad}G, \\
\\
\begin{array}{lllllll}
\displaystyle\quad\frac{{d\psi }}{dt}=u, &  & \displaystyle\frac{dT}{dt}=0,
&  & \displaystyle\frac{d\gamma }{dt}=-1, &  & \displaystyle\frac{dG}{dt}=0.%
\end{array}%
\end{array}%
\right.  \label{E}
\end{equation}%
Let us note: $\mathbf{C}=(\gamma +t)\func{grad}G$ and $d_{c}\mathbf{C}^{\ast
}=0$; the behaviour of $\varphi $ ($\displaystyle\frac{d\varphi }{dt}=%
\mathbf{V}^{2}$) is a consequence of the previous equations.

The two scalars $T$ and $G$ characterize the motion. The classification of
motions is in accordance with the respective positions of the surfaces $%
T=C^{ste} $ and $G=C^{ste}$ (when $T=s$ they are the isentropic and the Lamb
surfaces). We have different cases

(a) General case: $T$ and $G$ are independent (or $\func{grad}T\wedge\,
\func{grad} G\neq 0)$.

The $T$-surfaces (isentropic surfaces) and the $G$-surfaces (Lamb surfaces)
are different.

(b) $T$ is not constant and $G=G(T)$: the $T$-surfaces are $G$-surfaces.

Replacing $\psi $ par $\psi +\gamma\, G_{T}^{\prime }$ we get
\begin{equation}
\left\{
\begin{array}{c}
\mathbf{V}=\func{grad}\varphi +\psi \func{grad}\,T ,\\
\\
\begin{array}{ccc}
\displaystyle\quad\frac{{d\psi }}{dt}=u-G^{\prime }(T), &  & \displaystyle%
\frac{dT}{dt}=0.%
\end{array}%
\end{array}%
\right.  \label{F}
\end{equation}%
These equations characterize permanent oligotropic motions [6].

(c) $G$ is constant and $T$ is not constant.

This is a particular case of (b): ($\displaystyle\frac{{d\psi }}{dt}=u$).

(d) $T$ is constant, $G$ is not constant. The equations are
\begin{equation}
\left\{
\begin{array}{c}
\mathbf{V}=\func{grad}\varphi +\gamma\func{grad}\,G , \\
\\
\begin{array}{ccccc}
\displaystyle\quad\frac{d\gamma }{dt}=-1, &  & \displaystyle\frac{dG}{dt}=0,
&  & T=T_{o}.%
\end{array}%
\end{array}%
\right.
\end{equation}

\begin{figure}[tbp]
\begin{center}
\epsfig{file={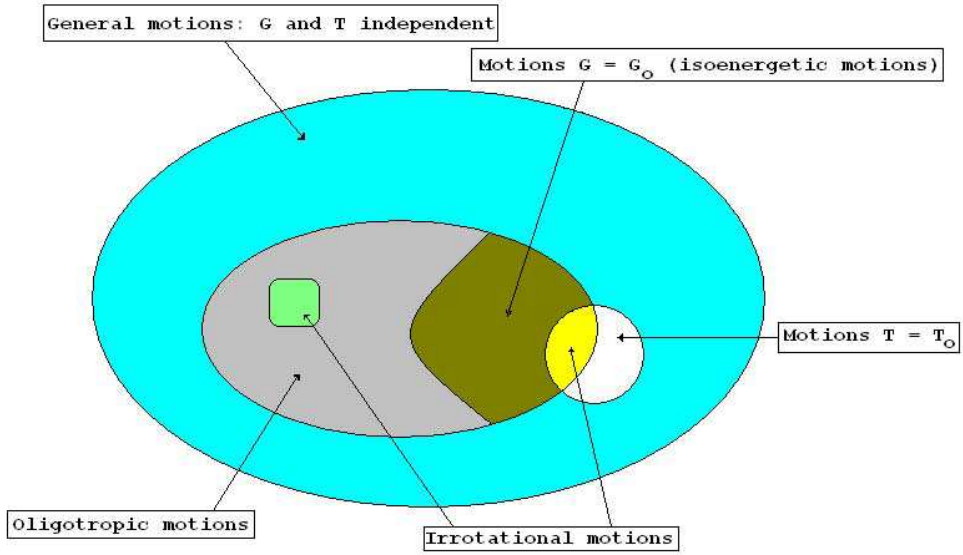}, width = 1\linewidth}
\end{center}
\caption{Classification of permanent motions: the different domains
do not mix along the motion of the fluid.} \label{fig2}
\end{figure}

(e) $T$ and $G$ are constant.

We obtain the irrotational motions
\begin{equation*}
\begin{array}{lllll}
\mathbf{V}=\func{grad}\varphi , &  & G=G_{o}, &  & T=T_{o}%
\end{array}%
\end{equation*}%
\emph{Let us note}: All the irrotational motions are not represented by case
(e) (such is the case of Hamel's motions [26]). In fact, relation (\ref%
{crocco}) points out that if the motion is irrotational and if one of the
mappings $T$ or $G$ is constant, the other mapping is also constant. If
neither $G$ nor $T$ are constant, $G$ must depend on $T$ (case (b)): an
irrotational motion is oligotropic.

Irrotational motions do not correspond to particular values of the
first integrals (or values convected by the flow). In Fig. 2, we
represent the flow of a fluid with different domains in
$\mathcal{D}_{t}$. The domains are convected by the flow and do not
overlap. If each domain has its own color, it keeps the color all
along the motion.

\section{Appendix}

The motion of the medium consists of the $t$-dependent $C^{2}$%
-diffeomorphism $\varphi _{t}\ $[5]
\begin{equation*}
\mathbf{X} \in \mathcal{D}_{o}\ \longrightarrow\ \mathbf{x} \in \mathcal{D}%
_{t}\,
\end{equation*}

We deduce $\displaystyle\frac{dF}{dt}=\displaystyle\frac{ \partial \mathbf{V}%
}{\partial \mathbf{x}}\, F$ and $\displaystyle\frac{dF^{-1}}{dt}=-F^{-1}%
\displaystyle\frac{\partial \mathbf{V}}{\partial \mathbf{x}}$.

Let us write $T^{\ast }(\mathcal{D}_{t})$ for the cotangent fiber bundle of $%
\mathcal{D}_{t}$ and $T_{\mathbf{x}}^{\ast }$($\mathcal{D}_{t})$ for the
cotangent linear space to $\mathcal{D}_{t}$ at $\mathbf{x}$; then,%
\begin{equation*}
\mathbf{x}\in \mathcal{D}_{t}\ \longrightarrow\ L(\mathbf{x},t)\in T_{%
\mathbf{x}}^{\ast }(\mathcal{D}_{t})
\end{equation*}%
represents a field of differential forms on $\mathcal{D}_{t}$. \newline
Let us write $T^{\ast }(\mathcal{D}_{o})$ the cotangent fiber bundle of $%
\mathcal{D}_{o}$ and $T_\mathbf{X}^{\ast }(\mathcal{D}_{o})$ the cotangent
linear space to $\mathcal{D}_{o}$ at $\mathbf{X}$. The mapping $\varphi
_{t}^{\ast }$ is induced by $\varphi _{t}$ for the form fields. The
convective derivation $d_{c}$ of a form field $L$ is deduced from the
diagram
\begin{equation}
\begin{array}{ccc}
L\in T^{\ast }(\mathcal{D}_{t}) &
\begin{array}{c}
\displaystyle\varphi _{t}^{-1\ast } \\
\longrightarrow \\
\\
\end{array}
& LF\in T^{\ast }(\mathcal{D}_{o}) \\
\begin{array}{c}
\displaystyle d_{c}\,\Big\downarrow \\
\end{array}
&  &
\begin{array}{c}
\displaystyle\Big\downarrow\, \displaystyle\frac{d}{dt} \\
\end{array}
\\
\begin{array}{c}
\\
\displaystyle\frac{dL}{dt}+L\displaystyle\frac{\partial \mathbf{V}}{\partial
x}%
\end{array}%
\ \ \ \ \ \  &
\begin{array}{c}
\varphi _{t}^{\ast } \\
\longleftarrow \\
\end{array}
&
\begin{array}{c}
\ \ \ \ \ \ \ \  \\
\ \ \ \displaystyle\frac{dL}{dt}F+L\displaystyle\frac{\partial \mathbf{V}}{%
\partial x}F\in T^{\ast }(\mathcal{D}_{o})%
\end{array}%
\end{array}
\label{graph}
\end{equation}
Let us note that
$\displaystyle\frac{dL}{dt}+L\,\displaystyle\frac{\partial
\mathbf{V}}{\partial \mathbf{x}}$ is the Lie derivative of $L$ with
respect to the velocity field $\mathbf{V}$ \footnote{Generally, for
any tensor, the convective derivative is the Lie derivative with
respect to the velocity field.}.
\newline The velocity field $\mathbf{V}$ is the infinitesimal
transformation of the one-parameter group of transformation $\varphi
_{t}$ [27].

\textit{Consequences} : $\mathcal{D}_{t}$ is assumed to be an
Euclidian space; let $b$ be a scalar field on $\mathcal{D}_{t}$. We
define two
form-fields as $\mathbf{V}^{\ast }$ and $(\func{grad}b)^{\ast }$. From (\ref%
{graph}) we deduce
\begin{equation*}
d_{c}(\mathbf{V}^{\ast })=\mathbf{\Gamma }^{\ast }+\frac{\partial }{\partial
x}\left( \frac{1}{2}\,\mathbf{V}^{2}\right)
\end{equation*}%
where $\mathbf{\Gamma }$ is the acceleration vector. Taking into account
that
\begin{equation*}
\begin{array}{ccccc}
d_{c}(\func{grad}b)^{\ast }=(\func{grad}\displaystyle\frac{db}{dt})^{\ast }
&  & \text{and} &  & d_{c}(LF^{-1})=\displaystyle\frac{dL}{dt}\,F^{-1}%
\end{array}%
,
\end{equation*}%
Eq. (\ref{motion1}) yields
\begin{equation}
d_{c}\mathbf{V}^{\ast }=\mathrm{grad}^{\ast }\left( \frac{1}{2}\mathbf{V}%
^{2}-g-\Omega \right) +\theta\, \mathrm{grad}^{\ast }s .  \label{20}
\end{equation}

\begin{lem}
Let $\theta $ and $s$ be two differential scalars fields in $\mathcal{D}_{t}$%
.\newline
There exist three scalar fields $\alpha $, $\beta $, $\gamma $ such that
\begin{equation}
\theta\, \mathrm{grad}^{\ast }s=\alpha\, \mathrm{grad}^{\ast }\beta +\mathrm{%
grad}^{\ast }\gamma  \tag{i}
\end{equation}%
is equivalent to: there exists a function $\varphi $ such that
\begin{equation}
\beta =\varphi (\theta ,s)  \tag{ii}
\end{equation}
\end{lem}

\emph{Proof}\,:

-\ If $\func{grad}\theta \wedge \func{grad}s=0$ $\ $then\ $\ \theta =\Phi
^{\prime }(s)$, $\theta \func{grad}s$ $=\func{grad}\Phi (s)$ \ and $\ \beta
=0$.

-\ If on an open set, $\func{grad}\theta \wedge \func{grad}s\neq 0$ \ then

\qquad $(i)\Rightarrow$\ $\func{grad}\theta \wedge \func{grad}s=\func{grad}%
\alpha \wedge \func{grad}\beta $. \newline
Consequently, $\func{grad}\theta ,\ \func{grad}s,\ \func{grad}\alpha $ and $%
\func{grad}\beta $ are in the same plane and there exist $\varphi $, $\psi $%
, $\chi $ such that $\beta =\varphi (\theta ,s)$, $\alpha =\psi
(\theta ,s)$  and $\gamma =\chi (\theta ,s)$. So, $(i)\Rightarrow
(ii)$.

$\qquad (ii)$    yields locally $
\theta $ $=h(\beta ,s)$. \newline
There exists $\gamma =H(\beta ,s)$ such that $H_{s}^{\prime }=\theta $. With
$\alpha =-H^{\prime }_\beta $, we obtain $\func{grad}^{\ast }\gamma
=H_{s}^{\prime }\func{grad}^{\ast }s+H_{\beta }^{\prime }\func{grad}^{\ast
}\beta $. So, $(ii)\Rightarrow (i)$.


\begin{lem}
Let $\sigma _{o}$ be an arbitrary non-constant differential scalar field in $%
\mathcal{D}_{o}$ independent of a given scalar field $\delta _{o}$ defined
on $\mathcal{D}_{o}$ by the relation $\delta _{o}(\mathbf{X})=\delta \lbrack
\varphi _{t}(\mathbf{X})]$. For any form field $\mathbf{X}\in \mathcal{D}%
_{o}\longrightarrow $ $\Xi _{o}^{\ast }(\mathbf{X})\in T_{X}^{\ast }(%
\mathcal{D}_{o})$ \, there exist -at least locally- three scalar fields
defined on $\mathcal{D}_{o}$ and denoted by $\alpha _{o}$,\, $\beta _{o}$
and $\omega _{o}$ such that
\begin{equation*}
\Xi _{o}^{\ast }=\mathrm{grad}_{o}^{\ast }\,\alpha _{o}+\beta _{o}\,\mathrm{%
grad} _{o}^{\ast }\,\delta _{o}+\omega _{o}\,\mathrm{grad}_{o}^{\ast
}\,\sigma _{o}
\end{equation*}
\end{lem}

Due to the fact that $\func{grad}_{o}^{\ast }\delta _{o}$ and $\func{grad}%
_{o}^{\ast }\sigma _{o}$ are two independent forms of $\mathcal{D}_{o}$, the
two surface-families $\delta _{o}(X)=a_{1}$ and $\sigma _{o}(X)=b_{1}$
(where $a_{1}$ and $b_{1}$ are two arbitrary constants), generate a curve
family denoted by $(\Gamma )_{a_{1}b_{1}}$. \newline
Let $(S)$ be any transverse surface of the curve family $(\Gamma
)_{a_{1}b_{1}}$, point $M_{o}$ denotes the current intersection between $%
(\Gamma )_{a_{1}b_{1}}$ and $(S)$. We denote by $\overset{{\Huge %
\curvearrowright }}{M_{o}M}$ the arc of the curve $(\Gamma )_{a_{1}b_{1}}$
with origin at $M_{o}$. \newline
We define a differential function as
\begin{equation*}
\mathbf{X}\in \mathcal{D}_{o}\longrightarrow \alpha _{o}(\mathbf{X})=\int_{%
\overset{{\huge \curvearrowright }}{M_{o}M}}\Xi _{o}^{\ast }\,d\mathbf{X}
\end{equation*}%
such that $d\delta _{o}=0$, $d\sigma _{o}=0$ and $d\alpha _{o}=\Xi
_{o}^{\ast }\,d\mathbf{X}$. Then, there exist -at least locally- two scalar
field forms $\beta _{o}$ and $\omega _{o}$ in $\mathcal{D}_{o}$ such that
\begin{equation*}
\Xi _{o}^{\ast }\,d\mathbf{X}-d\alpha _{o}=\beta _{o}\,d\delta _{o}+\omega
_{o}\,d\sigma _{o}.
\end{equation*}

\begin{thm}
For any mapping $\sigma $ of $\mathcal{D}_{t}$, independent of $\delta $ and
verifying $\displaystyle\frac{d\sigma }{dt} =0$, for any form field $\Xi
^{\ast }$ in $\mathcal{D}_{t}$ with a zero convective derivation, there
exist -at least locally- three mappings $\alpha $, $\beta $ and $\omega $ of
$\mathcal{D}_{t}$, verifying $\displaystyle\frac{d\alpha }{dt}=0$, $%
\displaystyle\frac{d\beta }{dt}=0$ and $\displaystyle\frac{d\omega }{dt}=0 $
such that $\Xi ^{\ast }=\func{grad}^{\ast }\alpha +\beta \func{grad}^{\ast
}\delta +\omega \func{grad}^{\ast }\sigma $.
\end{thm}

Let us consider $\sigma (t,\mathbf{x})=\sigma _{o}\,[\varphi _{t}^{-1}(%
\mathbf{X})]$, $\alpha (t,\mathbf{x}) $ $=\alpha _{o}\,[\varphi _{t}^{-1}(%
\mathbf{X})]$ , $\beta (t,\mathbf{x})=\beta _{o}[\varphi _{t}^{-1}(\mathbf{X}%
)]$ and the field $\Xi ^{\ast }$ of forms in $\mathcal{D}_{t}$ corresponding
to $\Xi _{o}^{\ast }$. \newline
Since $\sigma, \alpha, \beta, \omega $ and $\delta $ have zero material
derivatives, the previous forms and scalar functions have a zero convective
derivatives.

\textit{Consequences}:\newline
From the representation of the form field $\mathbf{C}^{\ast }$ and for $%
\delta =T$, we can get two scalar fields $\varphi $ and $\psi $ in System (%
\ref{equapot}); we obtain the potential representation (or Clebsch's
representation) for the flows in System (\ref{equapot1}). We note that the
choice of $\mathbf{C}^{\ast }$ is arbitrary and $\varphi $ and $\psi $ are
not unique.

\textit{Proof of system} (\ref{F}) :\newline
System (\ref{F}) is directly obtained when $G$ is independent of $T$. In the
converse, only a modification is needed: for a given form field $\Xi
_{o}^{\ast }$ in $\mathcal{D}_{o}$, there exist -at least locally- two
scalar fields $\alpha _{o}$ and $\beta _{o}$ such that
\begin{equation*}
\Xi _{o}^{\ast }=\mathrm{grad}_{o}^{\ast }\,\alpha _{o}+\beta _{o}\,\mathrm{%
grad}_{o}^{\ast }\,T_{o}
\end{equation*}%
It is an analog representation as Clesch's [16]. In this case, we consider a
transverse curve $(\Gamma )$ of $T$-surfaces and a scalar field
\begin{equation*}
\alpha _{o}(\mathbf{X})=\int_{\overset{{\huge \curvearrowright }}{M_{o}M}%
}\Xi _{o}^{\ast }\,d\mathbf{X}
\end{equation*}%
defined on $\mathcal{D}_{o}$ as the circulation of $\Xi _{o}^{\ast }$ on an
arbitrary curve on the $T$-surface connecting point $\mathbf{X}$ at point $%
\mathbf{X}_{o}$ where the surface intersects $(\Gamma )$. \newline
The circulation is independent of the curve connecting $\mathbf{X}_{o}$ and $%
\mathbf{X}$;  when $dT_{o}=0$ then $d\alpha _{o}=\Xi _{o}^{\ast
}\,d\textbf{X}$. Consequently there exists a Lagrangian multiplier
$\beta _{o}(\mathbf{X})$ such that
\begin{equation*}
\Xi _{o}^{\ast }\,d\mathbf{X}=d\alpha _{o}+\beta _{o}\,dT_{o}
\end{equation*}%
and the Clebsch form is obtained.

\end{document}